# Analyse scientométrique du domaine de l'infectiologie de 2000 à 2020

Lesya Baudoin[1]
lesya.baudoin@hceres.fr
Anne Glanard[1]
anne.glanard@hceres.fr
Abdelghani Maddi[1]
abdelghani.maddi@hceres.fr
Wilfriedo Mescheba[1]
wilfriedo.mescheba@hceres.fr
Frédérique Sachwald[1]
frederique.sachwald@hceres.fr

1. Observatoire des Sciences et Techniques, Hcéres, 2 Rue Albert Einstein, Paris, 75013 France

**Résumé**

Le domaine de l'infectiologie constitue un champ de recherche transversal. Afin de l'identifier précisément, l'article construit un corpus global des publications en infectiologie en combinant les moyens offerts par le langage contrôlé du thésaurus MeSH de Medline et par la catégorisation des revues scientifiques du Web of Science. Ce corpus mondial permet de caractériser les publications des 20 premiers pays publiant dans le domaine et de retracer les évolutions entre 2000 à 2020. La construction de cartes thématiques permet d'identifier les thèmes de recherche au sein de l'infectiologie dans le monde et en France. L'explosion des publications sur le Covid-19 en 2020 a un impact très visible sur la carte mondiale des thématiques en infectiologie et modifie le positionnement de certains pays dans ce domaine de recherche. La conclusion identifie des pistes d'approfondissement de l'analyse du domaine qui pourraient être suivies à mesure que des données plus complètes seront disponibles sur la période de la pandémie de Covid-19.

**Mots clés**

Publications scientifiques, Analyse scientométrique, Indice de spécialisation, Indice d'impact, France, Cartographie thématique, Infectiologie, Covid-19.

## Title: Scientometric Analysis of Research on Infectious diseases, 2000-2020

**Abstract**

Research on infectious diseases constitutes a transversal scientific field. A specific corpus is designed by combining a controlled language (Medline MeSH thesaurus) and the categorization of journals (Web of Science). From this global corpus, the article characterizes the publications from the top 20 countries publishing in the field and evolutions between 2000 and 2020. Topic maps show the research themes within the field of infectious diseases both in the world and in France. The explosion of publications on Covid-19 in 2020 has a quite visible impact on the topic map in infectious diseases and changes the position of some countries in this field of research. The conclusion points to issues for further research as more complete data will become available on the Covid-19 period.

**Keywords**

Scientific publications, scientometric analysis, specialisation index, impact index, France, topic mapping, infectious diseases, Covid-19

# Introduction

L'infectiologie constitue un domaine particulièrement intéressant à étudier avec les outils de la scientométrie au moment où la recherche y connaît une croissance extraordinaire du fait de la pandémie de Covid-19. Cet article construit un corpus mondial des publications en infectiologie sur la période 2000-2020 afin de mesurer la croissance des publications dans le domaine de l'infectiologie, d'analyser sa structure thématique interne et de comparer les principaux pays producteurs à partir de différents indicateurs.

Le cas de la France est plus particulièrement analysé. De Louis Pasteur aux découvreurs du VIH, la recherche française en infectiologie compte de nombreux chercheurs dont l'apport scientifique a été remarquable. Dotée d'institutions renommées renforcées par des investissements à travers des Labex et Equipex, ainsi que de réseaux de la recherche hospitalière développés, la France possède d'importantes ressources scientifiques dans le domaine. Ses publications dans le champ des recherches sur l'immunité et l'infection représentent une part relativement importante du total de ses publications scientifiques en comparaison de la part de ce domaine dans les publications mondiales (OST 2021). Il apparaît donc intéressant de comparer l'évolution des publications françaises à celles d'autres pays en réponse à la pandémie de Covid-19.

L'infectiologie est un domaine de recherche transversal : il se nourrit des recherches en microbiologie, virologie, immunologie et génétique. Dans les domaines cliniques, au-delà de la spécialité « maladies infectieuses et tropicales », l'infectiologie est notamment très présente dans les soins critiques, la chirurgie et l'oncologie. La pharmacologie et la vaccinologie sont également étroitement liées à l'infectiologie. La constitution d'un corpus du domaine de l'infectiologie présente ainsi un enjeu méthodologique qui est abordé dans la première partie consacrée aux données et aux méthodes déployées dans l'article. La deuxième partie cartographie les principales thématiques présentes au sein du domaine de l'infectiologie et leur évolution en 2020. Le profil thématique du corpus des publications françaises est plus particulièrement analysé. La troisième partie compare la dynamique et les caractéristiques des publications en infectiologie des principaux pays producteurs depuis 2000. La conclusion revient sur les principaux résultats et sur les approfondissements qui paraissent utiles.

# Données et méthode

## Données

Le corpus est constitué en combinant deux sources de données. La première source, PubMed/Medline, met en œuvre un vocabulaire contrôlé organisé, le thésaurus MeSH (Medical Subject Headings). Les termes MeSH sont des mots-clés normalisés liés par des relations sémantiques ; ils sont attribués manuellement par les indexeurs pour décrire le contenu des articles[1]. Les descripteurs MeSH sont organisés en 16 catégories[2], chacune étant divisée en sous-catégories. Au sein de chaque sous-catégorie, les descripteurs sont organisés hiérarchiquement du plus général au plus spécifique jusqu'à treize niveaux. Le moteur de recherche de PubMed dispose d'une fonctionnalité qui identifie automatiquement tous les termes situés plus bas dans la hiérarchie MeSH. Cette fonctionnalité, appelée « explosion », permet d'optimiser les requêtes. Il est cependant nécessaire de s'assurer que tous les termes plus fins entrent dans le champ d'analyse. Par ailleurs, la recherche PubMed permet de pondérer les termes MeSH. Elle distingue les sujets majeurs de l'article, généralement obtenus à partir du titre et/ou de l'énoncé des objectifs, des sujets ayant une moindre importance dans l'article.

La seconde source de données mobilisée est la base de publications OST-WoS, version enrichie du Web of Science de Clarivate Analytics. Cette base de données indexe les publications scientifiques et leurs citations à l’échelle mondiale ; il s'agit de la source de données la plus ancienne et l’une des plus utilisées en scientométrie. Les plus de 20 000 revues indexées sont classées en 254 spécialités (WoS categories), qui, comme toutes les classifications basées sur des revues, n’est pas particulièrement spécifique. En outre, cette base ne dispose pas d’un vocabulaire normalisé et doit être interrogée en langage naturel, ce qui rend difficile la constitution de corpus thématiques précis.

[1] Voir, https://www.adbs.fr/langages-documentaires#th%C3%A9saurus

[2] https://meshb.nlm.nih.gov/treeView

## Constitution du corpus

Le domaine de l'infectiologie étant transversal, la construction d'un corpus de publications représentatif demande une méthode adaptée. Un équilibre doit être trouvé entre couverture du domaine d'une part et précision ou pertinence d'autre part. La méthode retenue ici combine les avantages des deux sources de données mobilisées. Les articles pertinents ont été extraits de Medline en utilisant l’équation de recherche (1).

Equation (1) :

("Infections"[MH] OR "Bacterial Vaccines"[MH] OR "Fungal Vaccines"[MH] OR "Protozoan Vaccines"[MH] OR "Toxoids"[MH] OR "Viral Vaccines"[MH] OR "Disease Notification"[MH] OR "Disease Eradication"[MH] OR "Disease Transmission, Infectious"[MH] OR "Contact Tracing"[MH] OR "Carrier State"[MH] OR "Chain of Infection"[MH] OR "Disease Outbreaks"[MH] OR "Travel-Related Illness"[Mesh:NoExp] OR "Quarantine"[MH] OR "Reinfection"[MH]) AND 2000/01/01:2020/12/01[dp]

Les articles identifiés ont ensuite été mis en correspondance[3] avec la base OST-WoS dans sa version d’avril 2021. Ce premier corpus issu de PubMed a été complété en ajoutant les articles des revues des catégories WoS INFECTIOUS DISEASES et TROPICAL MEDICINE.

Le repérage des publications est effectué sur l’ensemble de la base OST-WoS, c’est-à-dire pour tous les types de documents sur le périmètre des index suivants du WoS : SCI-Science Citation Index Expanded, SSCI-Social Sciences Citation Index, A&HCI-Arts & Humanities Citation Index, CPCI-Conference Proceedings Citation Index (S et SSH). L’analyse est ensuite réalisée en ne retenant que les contributions scientifiques représentées par les types de publication suivants : les articles originaux (Article) y compris ceux des actes de colloques (Proceedings Paper), les synthèses (Reviews) et les lettres (Letters). Les documents pour lesquels manque une partie des métadonnées (adresse, catégorie WoS) ne sont pas pris en compte ; les publications rétractées sont également exclues.

L’année de publication la plus récente disponible dans la version d’avril 2021 de la base est 2020, date pour laquelle les données ne sont pas tout à fait complètes. De ce fait, les indicateurs pour 2020 ne sont pas définitifs. En plus du nombre de publications, deux indicateurs sont calculés : l’indice de spécialisation et l’indice d’impact normalisé[4]. Le premier mesure l’activité des pays dans une discipline/thématique donnée ; il est obtenu en rapportant la part de la discipline dans l’ensemble des publications du pays à la même part au niveau du monde. Le deuxième mesure l’intérêt suscité par les publications de l’acteur au sein de la communauté scientifique. L’impact normalisé par domaine de recherche d’un pays est défini par le nombre moyen de citations des publications du pays d’une année donnée, normalisé par la moyenne mondiale des citations obtenues par les publications de ce domaine la même année. Ces indicateurs sont indépendants de la taille des pays et permettent des comparaisons internationales.

[3] L’intégration récente des identifiants PMID de PubMed dans la base WoS facilite l’utilisation conjointe des deux bases. Cependant l’appariement entre identifiants PubMed et WoS n’est pas parfait et demande d’effectuer des contrôles.

[4] L’indice d’impact normalisé est basé sur le nombre de citations de chaque article et non sur la moyenne des citations d’une revue, comme le facteur d’impact.

## Description générale du corpus

L'équation de recherche (1) a permis d'identifier 1 495 000 publications parues entre 2000 et 2020. Sur l'ensemble de ces publications issues de PubMed, 16,6% sont absentes de la base OST-WoS et 18,6% correspondent aux publications des catégories WoS INFECTIOUS DISEASES et TROPICAL MEDICINE. Symétriquement, 39,2% des publications extraites de la base OST-WoS à partir des catégories INFECTIOUS DISEASES et TROPICAL MEDICINE ne figurent pas dans les publications obtenues à partir de PubMed.

**Tableau 1 : Nombre de publications par origine des données**

| Origine des données | Nombre de publications identifiées |
|---|---|
| Publications issues de PubMed présentes dans la base OST-WoS | 968 055 |
| Publications issues de PubMed absentes dans la base OST-WoS | 248 578 |
| Publications issues de PubMed présentes dans liste des publications des 2 spécialités WoS | 278 367 |
| Publications des 2 spécialités WoS absentes de la liste issue de PubMed | 179 099 |

*Source : Medline, base OST, Web of Science, traitement OST*

Les publications absentes de la base OST-WoS ne sont pas prises en compte. Le corpus est ainsi réduit à 1 426 270 publications. L'application des filtres indiqués précédemment (présence d'adresse d'affiliation et de spécialiate WoS) réduit le corpus « Infectiologie » à 1 215 199 publications. Ce corpus est mobilisé dans la partie consacrée au positionnement des principaux pays publiant dans le domaine (voir plus bas).

L'étude de la cartographie des thématiques au sein du domaine de l'infectiologie nécessite des mots clés auteurs, ce qui limite les publications pouvant être retenues pour l'analyse. En effet, sur l'ensemble des publications identifiées et présentes dans le périmètre OST, 732 593 (60%) disposent de mots clés auteurs. Ces mots clefs sont traités à l'aide d'outils d'analyse textuelle afin d'harmoniser les orthographes et regrouper ceux qui portent la même racine (ex : pluriel et singulier). Après le nettoyage (publications sans mots-clés ou qui présentent de faux mots-clés (ex. caractères spéciaux) etc. , 3 447 publications ont été supprimées. Le sous-corpus « Infectiologie cartographie » comporte 729 146 publications parues entre 2000 et 2020.

Par ailleurs, un sous-corpus Covid-19 a été constitué à l'aide de 32 mots clés[5] qui peuvent être résumés en 6 mots : « 2019 ncov », « coronavirus », « Covid », « sars cov 2 », « wuhan seafood market pneumonia virus » et « sars cov2 ». Cette sélection s'appuie sur l'hypothèse que les publications parues entre 2019 et 2020 font référence à la pandémie en cours. L'application de ces filtres a permis d'identifier 19 515 publications parues en 2019-2020 et comportant des mots clés auteurs.

# Cartes thématiques des publications en infectiologie dans le monde et en France

Un des enjeux majeurs de l'analyse scientométrique est la compréhension de l'évolution des sciences et la formation des spécialités disciplinaires. Cet enjeu est d'autant plus central que la science évolue de façon spectaculaire donnant naissance à de nouveaux objets de recherche, au point qu'il est désormais impossible pour un chercheur d'appréhender l'ensemble de la littérature au sein même de sa spécialité.

[5] Voir liste des mots clefs en annexe (Tableau A1).

Vers la fin des années 2000, plusieurs outils permettant d'analyser les liens entre les publications scientifiques ont été développés, aussi bien au niveau thématique qu'institutionnel. Ces outils fournissent des représentations visuelles des réseaux des auteurs ou des termes utilisés dans les publications scientifiques (dans les titres résumés ou mots-clés).

L'analyse thématique du corpus « Infectiologie cartographie » utilise le logiciel VOSviewer qui permet de générer des cartes à partir de données bibliographiques ou textuelles. La construction de la carte mondiale a suivi les 4 grandes étapes d'une cartographie statique : indexation des termes, construction de la matrice de cooccurrence, mesure de la similarité et constitution des clusters (Chavalarias et al, 2013 ; Van Eck et al, 2014). La représentation des termes sur une carte dans un espace à deux dimensions tient compte de la proximité entre les termes d'un point de vue de leur cooccurrence. Ainsi, plus deux termes reviennent ensemble régulièrement dans la liste des mots clés des publications, plus ils seront proches sur la carte et feront partie d'un même cluster. La taille des bulles des termes est représentative de leur nombre d'occurrences dans le corpus.

## Carte mondiale des thématiques en infectiologie

La carte mondiale (Fig. 1) est structurée en 7 clusters thématiques décrits par le tableau 2. Les thèmes identifiés tiennent compte de l'ensemble des termes sous-jacents et pas uniquement de ceux qui sont visibles sur l'image.

**Fig.1. Cartographie thématique du corpus infectiologie, monde**

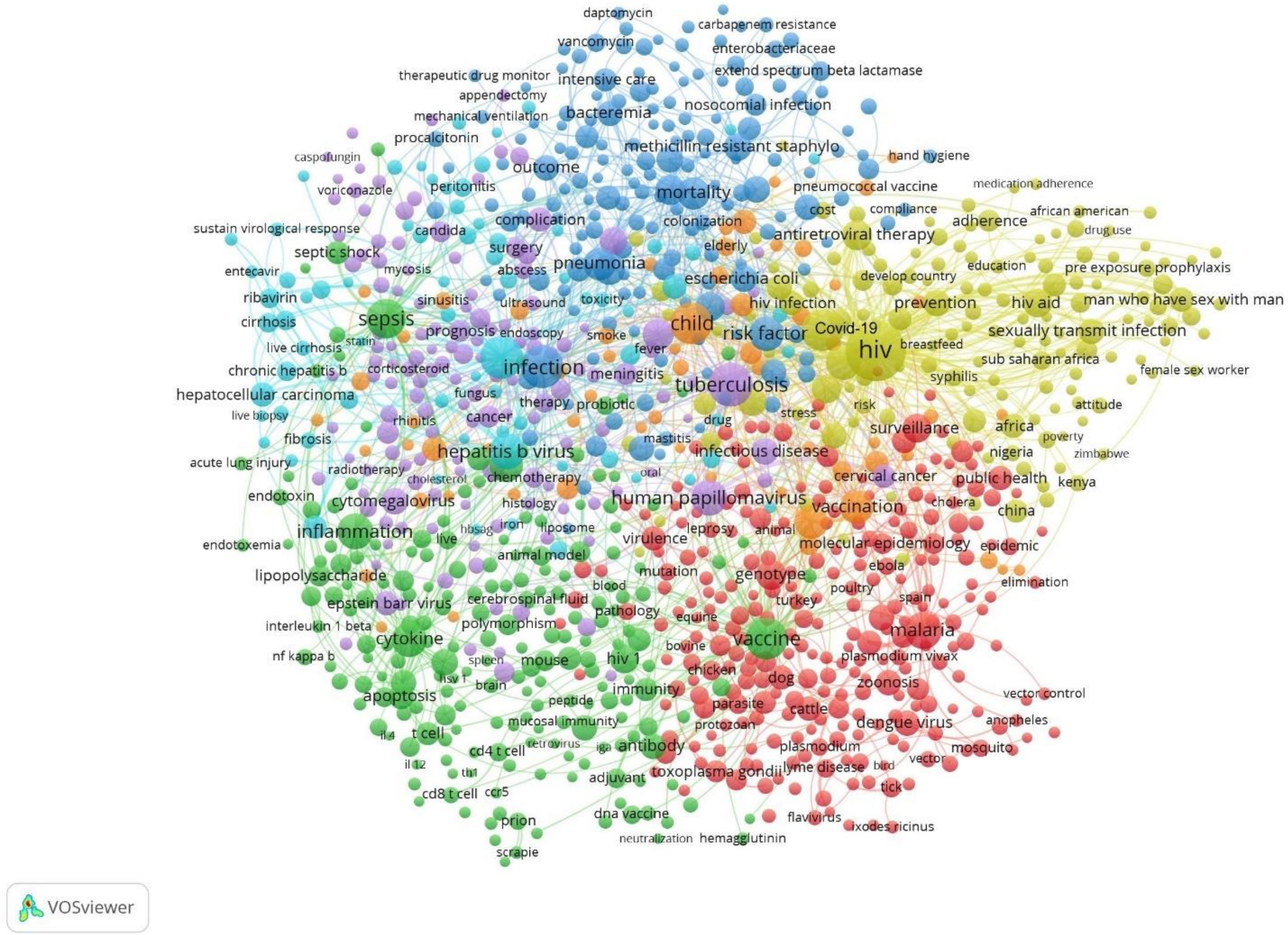

*Source : Base OST, Web of Science, traitement OST*

La carte mondiale présente une opposition sur l’axe horizontal d’ouest en est entre les approches de biologie fondamentale (clusters 2 et 6) et d’épidémiologie-santé publique (clusters 1 et 4). L’axe vertical présente au nord les infections à transmission directe (clusters 3, 4) et au sud les infections à transmission vectorielle (cluster 1). Les frontières entre les clusters ne sont pas étanches : la délimitation des clusters est basée sur les propriétés statistiques, ce qui n’exclut pas les liens existants entre les mots des différents clusters. Ainsi, certains mots-clefs appartenant à un cluster donné peuvent néanmoins être fortement liés à d’autres clusters. Par exemple, « Sepsis » appartient statistiquement au cluster 2, alors qu’il apparaît sur la carte à la fois excentré et très lié aux clusters 5 et 6.

**Tableau 2 : Description des clusters de la carte thématique mondiale**

| Numéro du cluster | Titre | Composition |
|---|---|---|
| 1 | Zoonoses, maladies à transmission vectorielle, maladies tropicales | - Vecteurs animaux<br>- Modélisation des épidémies<br>- Études séro-épidémiologiques<br>- Épidémiosurveillance<br>- Aspects environnementaux des épidémies<br>- Phylogéographie des agents infectieux<br>- Génomique et génétique des populations des microorganismes pathogènes |
| 2 | Mécanismes fondamentaux de l’infection et mécanismes moléculaires et cellulaires impliqués dans l’interaction hôte-pathogène | - Biologie cellulaire de l’infection<br>- Régulation de la transcription par l’hôte<br>- Réponse immunitaire, inflammation et sepsie<br>- Réponse métabolique<br>- Immunomodulation<br>- Immunité innée |
| 3 | Infections d’origine bactérienne et résistance aux agents antibactériens<br>Infectiologie en milieu hospitalier : soins critiques, infections postopératoires, infections sur cathéters, infections nosocomiales | - Prévention des infections<br>- Pharmacologie des médicaments antibactériens<br>- Facteurs de virulence bactérienne<br>- Antibio-résistance |
| 4 | Maladies infectieuses à transmission directe et santé publique | - VIH, Covid-19 : aspects épidémiologiques<br>- IST : Facteurs de risque et populations à risque<br>- Complications infectieuses de la grossesse et transmission materno-foetale de l'infection<br>- Bien-être et santé mentale des patients<br>- Facteurs socio-économiques<br>- Mesures préventives et prise en charge des patients<br>- Economie de la santé liée aux maladies infectieuses |
| 5 | Infections opportunistes et secondaires | - Infections chez les patients immunodéprimés<br>- Infections opportunistes : infections d’origine fongique, tuberculose, HPV<br>- Infection comme suite des traitements immunomodulateurs et immunosuppresseurs |
| 6 | Hépatites virales et traitements antiviraux | Hépatites virales, traitements antiviraux. Comorbidités. |
| 7 | Maladies infectieuses communes | - Infections respiratoires aéroportées<br>- Maladies transmissibles de l’enfant<br>- Politiques vaccinales |

Une fois la carte mondiale construite, il est possible d’analyser la présence des différents termes qui la composent à travers le temps en calculant les scores de concentration par année. La carte présentée en Annexe (A1) permet de visualiser l’évolution temporelle des thématiques de recherche. Cette carte présente un focus sur des liens de cooccurrence du terme « Covid-19 » et signale la concentration des différents termes par année par un dégradé de couleurs. Ainsi, par exemple le terme « SARS » (Severe Acute Respiratory Syndrome Coronavirus) est très présent dans les années antérieures à 2010, alors que « Covid-19 » est concentré sur 2020.

## Profil thématique de la France en infectiologie

La carte thématique des publications contenant au moins une adresse française a été superposée sur la carte mondiale afin de positionner les thématiques de la recherche française en infectiologie. Des indices d’activité par terme permettent d’identifier les thématiques où la recherche française connait une activité relativement intense. Dans un premier temps, le poids relatif de chaque terme est calculé dans la carte de la France et dans celle du monde. Dans un second temps, le poids du terme pour la France est rapporté à son poids dans le monde. La figure 2 cartographie les thématiques de spécialisation de la France à partir de ces indices relatifs. Les termes présents sur la carte signalent des liens de cooccurrence au moins 50% plus élevés pour les publications françaises que pour l'ensemble mondial. Par exemple, le terme « *africa* » est 2,5 fois plus présent dans les mots-clés des publications françaises que dans celles du monde. Cela signifie que l'Afrique occupe une place beaucoup plus importante dans les publications de la France en infectiologie que dans l'ensemble des publications mondiales.

**Fig.2. Cartographie des thématiques de spécialisation de la France en infectiologie, 2000-2020**

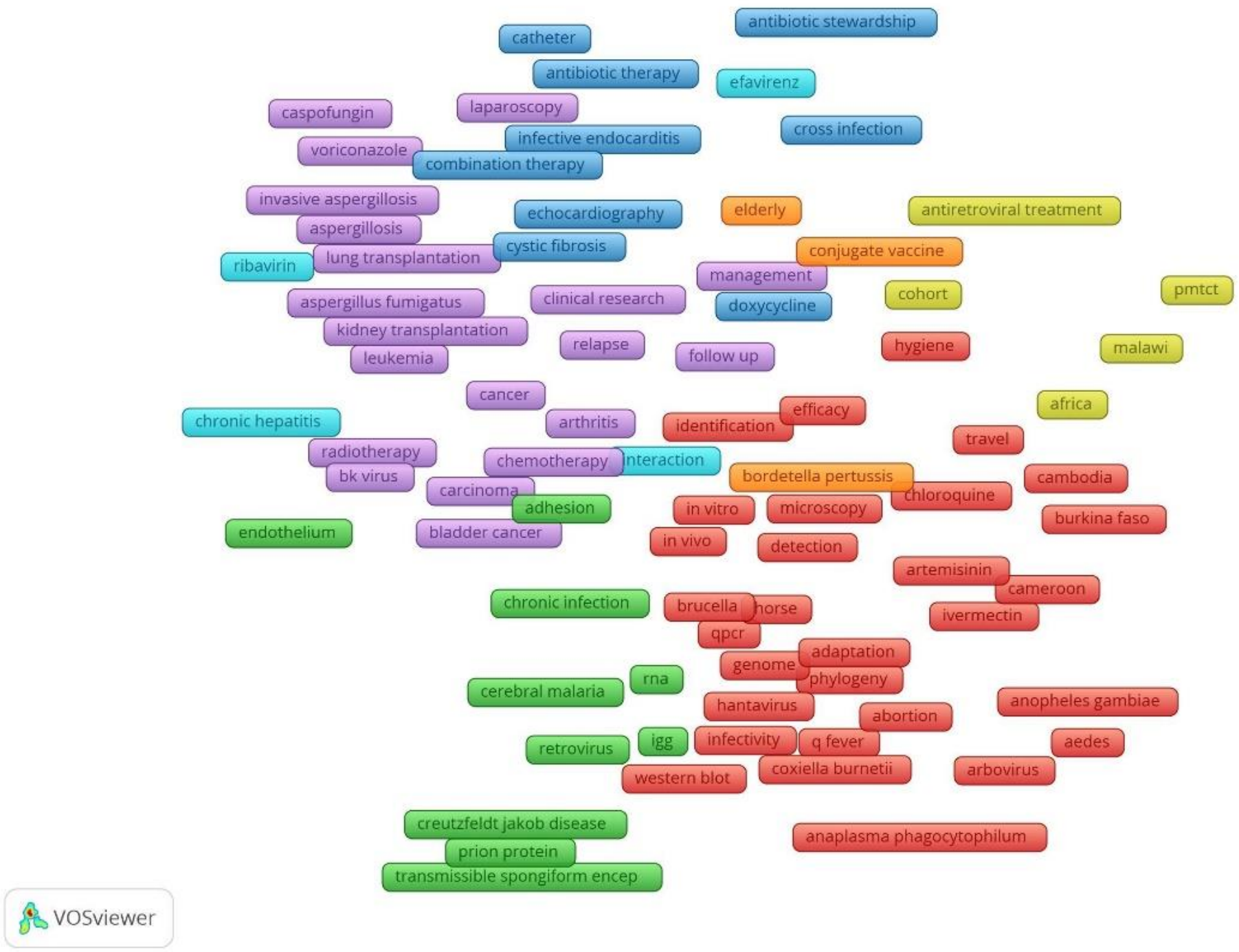

*Source : Base OST, Web of Science, traitement OST*

Globalement, la différence observée entre les thématiques Monde et France se traduit par une concentration des sujets de spécialisation française dans les clusters 1 et 5. Inversement, les sujets de spécialisation de la France sont moins concentrés dans les clusters 2 et 4. Les thèmes dont la présence dans le corpus des publications françaises est relativement plus importante que dans le corpus mondial sont présentés dans la liste suivante organisée par cluster :

- ***Cluster 1 :*** études génétiques et phylogénétiques des agents infectieux ; approches écologiques et évolutives de la lutte contre les infections à transmission vectorielle et les zoonoses.
- ***Cluster 2 :*** étude clinique et moléculaire des rétrovirus ; maladies à prions ; infection chronique ; choc septique.
- ***Cluster 3 :*** résistance aux antibiotiques et antibiogouvernance ; risques infectieux liés aux soins intensifs.
- ***Cluster 4 :*** transmission du VIH en Afrique, transmission materno-fœtale, études des cohortes VIH.
- ***Cluster 5 :*** recherche clinique en infectiologie des maladies opportunistes, chez les patients immunodéprimés ; infections fongiques et traitements antifongiques.
- ***Cluster 6 :*** hépatites ; traitements antiviraux.
- ***Cluster 7 :*** vaccination contre les infections rhinopharyngées d'origine bactérienne, infections respiratoires chez les personnes âgées.

# Caractérisation des publications en infectiologie des principaux pays producteurs

## Des dynamiques nationales variées

Le nombre de publications en infectiologie a été multiplié par plus de deux en vingt ans, avec une progression moindre que celle de l'ensemble des publications scientifiques entre 2000 et 2019[6] (Figure 3). L'année 2020 constitue donc une rupture : le nombre de publications mondiales en infectiologie fait un bond pour atteindre 118 148, alors que le recueil des publications parues en 2020 est encore incomplet. Entre 2019 et 2020, le nombre de publications en infectiologie a progressé de près de 40%, ce qui représente une augmentation spectaculaire dans la base de données pour un domaine de recherche. Les publications dédiées au Covid-19 contribuent fortement à cette augmentation, mais ne l'expliquent pas à elles toutes seules (Figure 3).

**Fig.3. Évolution du nombre de publications en infectiologie dans le monde, 2000-20***

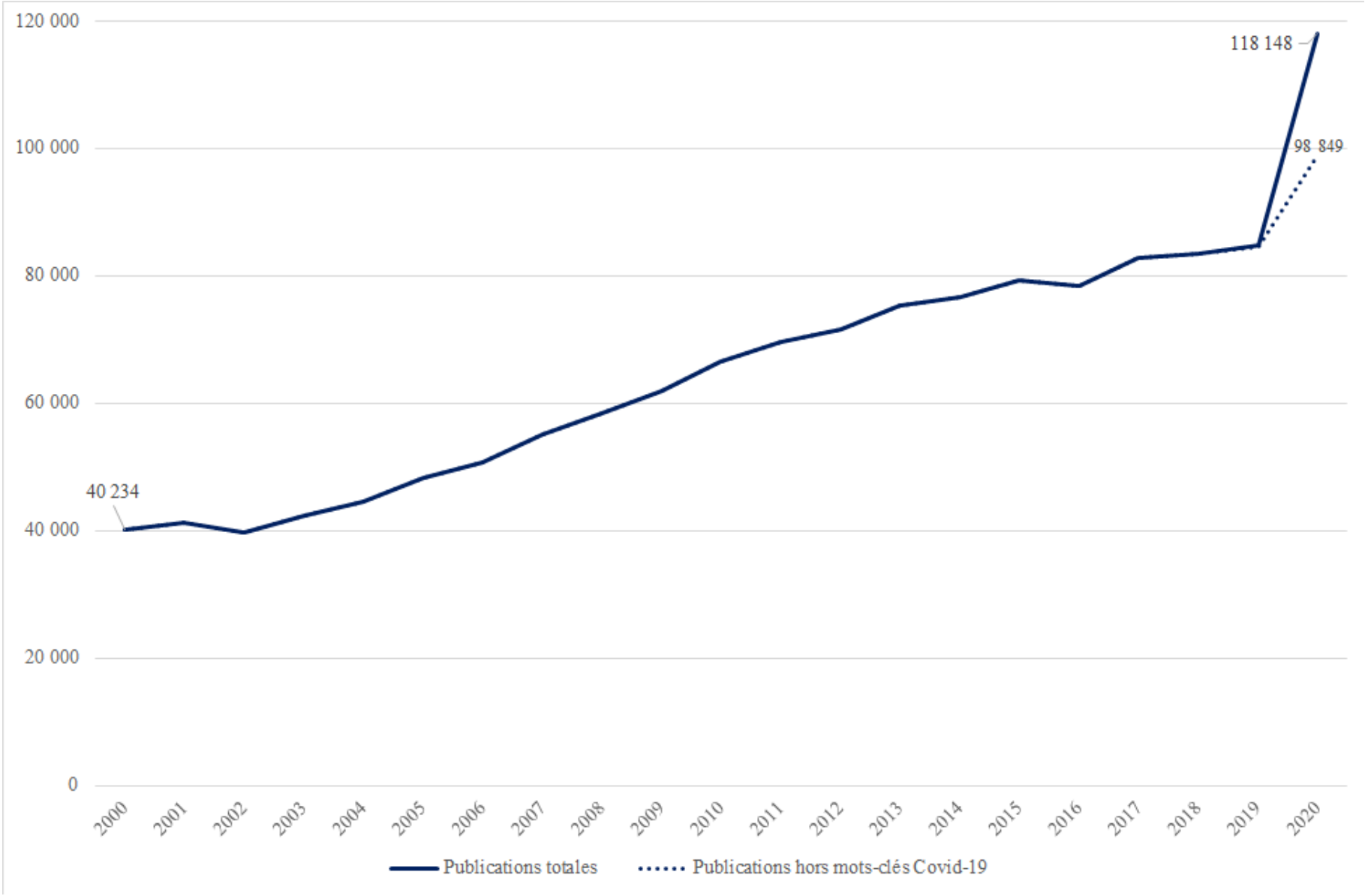

* année 2020 incomplète.

*Source : Base OST, Web of Science, traitements OST*

Depuis 2000, les positions nationales évoluent sensiblement pour certains pays (figure 4). Les publications de la Chine augmentent rapidement depuis 2010 et elle devient le second pays publiant le plus en 2013. Les États-Unis conservent leur position de leader, le Royaume-Uni passant en 3ème position. La France, en 3ème pays publiant le plus en 2000 est devancée par la Chine en 2010, par le Brésil en 2017, puis en 2020 par l'Italie et l'Inde. Le Brésil a fortement augmenté ses publications depuis les années 2000 et est passé de la 11ème position en 2000 à 4ème en 2017. L'Italie quant à elle enregistre une très forte progression en 2020 avec un nombre de publications qui double.

[6] La croissance du total des publications de 2000 à 2019 a été de 136%, contre 111% pour l'infectiologie

**Fig.4. Nombre de publications des 20 premiers producteurs en infectiologie, 2000-20**

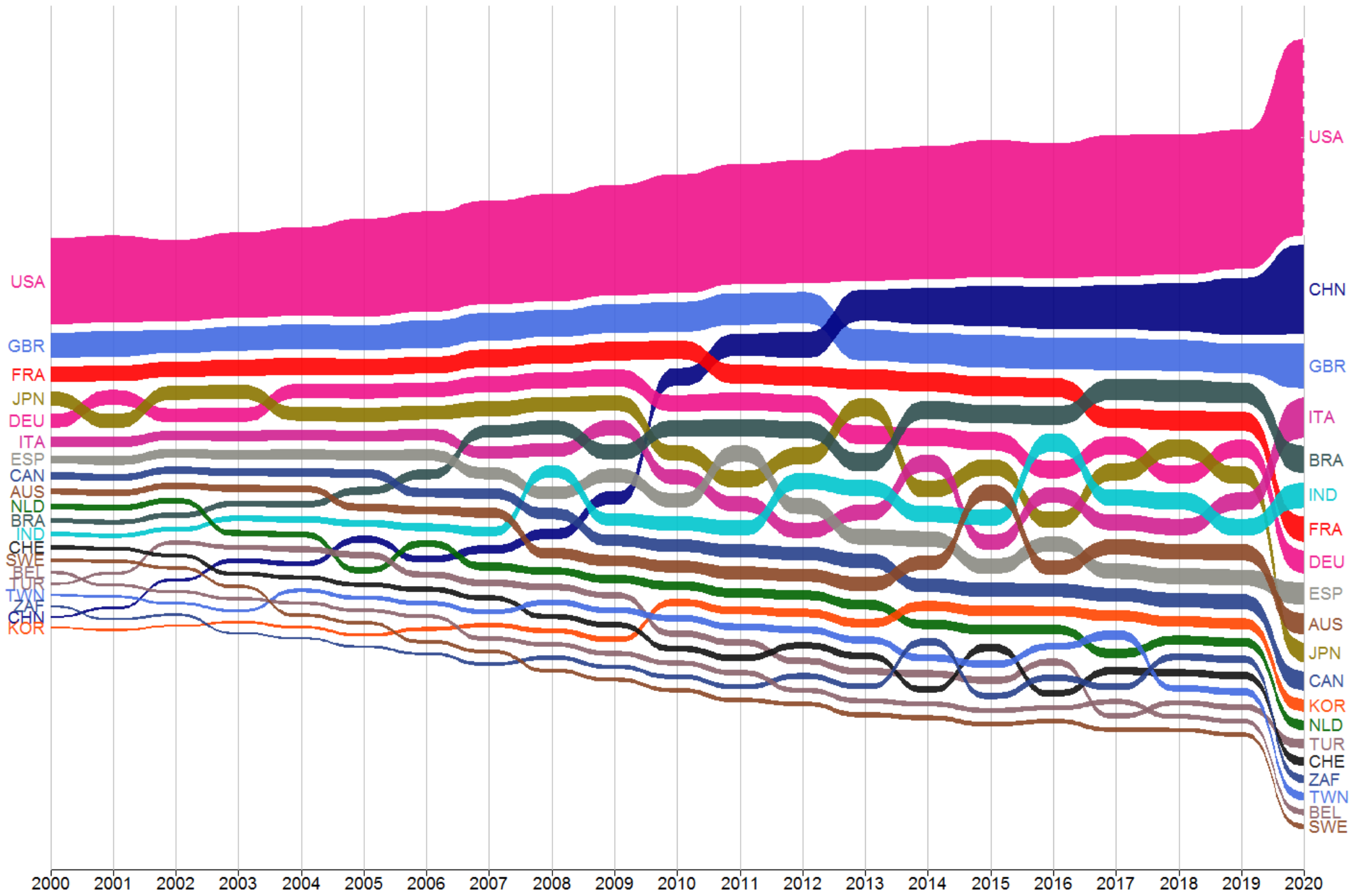

*Source : Base OST, Web of Science, calculs OST*

Au total, la figure 4 suggère que l'explosion des publications en infectiologie en 2020 due à l'émergence de la pandémie de Covid résulte d'une mobilisation variable des systèmes de recherche sur ce thème émergent.

## Évolution de la spécialisation des principaux producteurs

Le profil thématique des pays au sein du domaine de l'infectiologie contribue à expliquer la mobilisation plus ou moins forte et rapide des systèmes de recherche sur le nouveau sujet qu'a été le Covid à partir de 2020.

L'indice de spécialisation scientifique rapporte la part d'un domaine dans le total des publications d'un pays à ce même ratio pour le monde. La valeur neutre de l'indice de spécialisation est 1 ; lorsque l'indice est supérieur à 1, le pays est spécialisé dans le domaine par rapport à la référence.

Le pays le plus spécialisé en infectiologie parmi les principaux producteurs est l'Afrique du Sud (Figure 5). Son nombre de publications est modeste, mais la part de l'infectiologie dans ses publications est près de 3 fois supérieure à la part du domaine dans les publications mondiales en 2015-19. Cette forte spécialisation s'explique notamment par l'importance des recherches menées en Afrique du Sud sur le Sida et la tuberculose. Le Brésil est le 2$^{ème}$ pays le plus spécialisé dans le domaine. La Chine augmente sensiblement son degré de spécialisation sur la période, tout en conservant un indice inférieur à la moyenne mondiale.

**Fig.5. Indice de spécialisation en infectiologie des 20 premiers pays publiant le plus dans le domaine**

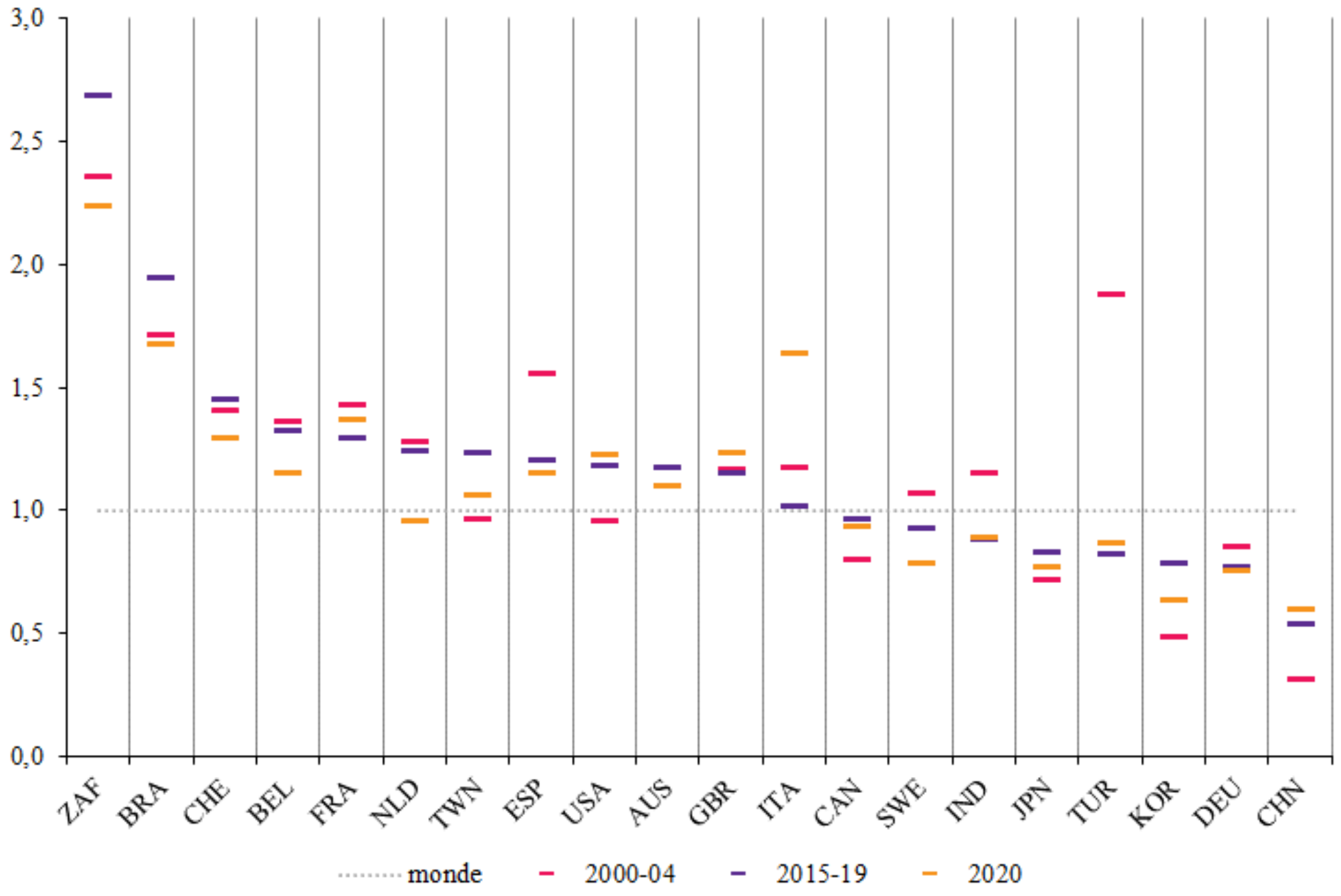

*Source : Base OST, Web of Science, calculs OST*

Le degré de spécialisation de la France en infectiologie est assez élevé (1,3) parmi les pays à revenus élevés : proche de celui de la Suisse, de la Belgique et des Pays Bas sur l'ensemble de la période. La France est ainsi plus spécialisée en infectiologie que les Etats-Unis, le Royaume Uni ou l'Allemagne (non spécialisée). Les pays à hauts revenus, souvent les plus spécialisés en biologie fondamentale et en recherche médicale (OST 2021) ne sont donc pas toujours spécialisés en infectiologie.

Plusieurs pays augmentent légèrement leur spécialisation en infectiologie en 2020 par rapport à 2015-19 : la Chine, le Royaume Uni ou la France notamment. L'Italie connaît une augmentation beaucoup plus forte de sa spécialisation, avec un indice qui croît de 60% pour dépasser 1,6. Cette évolution est liée au doublement du nombre de publications italiennes en infectiologie (Figure 4).

La figure 6 souligne la mobilisation rapide de certains pays sur le thème du Covid-19. Le grand nombre de publications de l'Italie dès le début de la pandémie se lit dans l'indice de spécialisation sur le thème du Covid-19, près de deux fois plus élevé que son indice de spécialisation en infectiologie. A l'inverse, la recherche en Afrique du Sud, très spécialisée en infectiologie, n'apparaît pas s'être mobilisée sur le Covid-19.

Le cas de l'Italie peut s'expliquer en partie par le fait que le pays a été durement touché dès le début de la pandémie. A ce stade précoce, les nombreuses publications portaient surtout sur des observations cliniques et la prise en charge des malades (Odone et al. 2020, Turatto et al. 2021). Outre l'Italie, la Turquie et la Chine sont plus spécialisés sur le thème du Covid qu'en infectiologie en général.

**Fig.6. Indice de spécialisation en recherche sur le Covid-19 des 20 premiers producteurs, 2020***

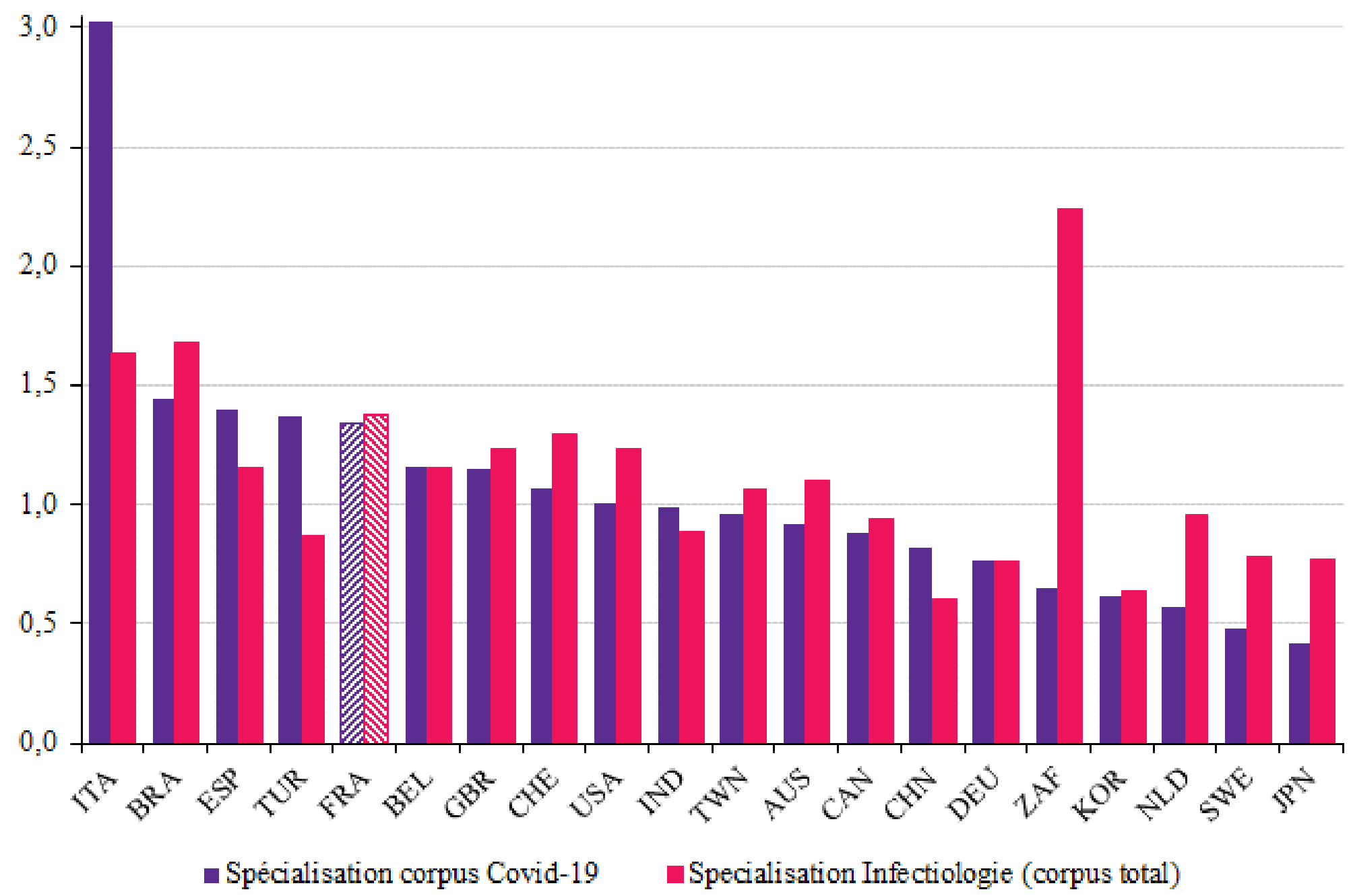

*Source : Base OST, Web of Science, calculs OST*

La France n'est pas plus spécialisée sur le thème du Covid que sur l'ensemble du domaine de l'infectiologie. Sur la première année de la pandémie, les données de publications disponibles suggèrent que la recherche française s'est pas mobilisée au-delà de sa structure déjà spécialisée en infectiologie. Ce constat très partiel à ce stade pourrait être dû en partie au fait que la France a été touchée plus tard que l'Italie, la Chine ou l'Espagne et que les premières descriptions cliniques avaient déjà été publiées. Les analyses ultérieures ont demandé la mise en place d'études plus approfondies et le recrutement de patients. Or, de ce point de vue, si la France a rapidement débloqué des financements, la coordination des recherches a été insuffisante, ce qui a pu allonger les délais, voire menacer l'aboutissement de certains travaux (IGESR 2021, Cour des comptes 2021).

## Impact scientifique des publications des principaux pays producteurs

L'impact normalisé par domaine de recherche d'un pays est défini par le nombre moyen de citations des publications du pays d'une année donnée, normalisé par la moyenne mondiale des citations obtenues par les publications de ce domaine la même année. L'indice est normalisé : il est calculé au niveau de chaque spécialité composant les disciplines afin de tenir compte de la structure disciplinaire des publications des pays. La valeur neutre de l'indice est 1 et un indice supérieur à 1 signifie que les publications du pays ont un impact supérieur à la moyenne mondiale.

La figure 7 fournit les indices d'impact jusqu'à l'année 2019 avec une fenêtre de citation de 2 ans. Les publications sur le Covid-19 ne peuvent donc pas être prises en compte pour cet indicateur. La figure 7 indique qu'en 2015-19, la Suisse, les Pays Bas, les Etats-Unis, la Suède, l'Allemagne, l'Australie et le Royaume Uni ont des indices au moins 20% au-dessus de la moyenne mondiale.  Ces mêmes pays ont aussi des indices d'impact parmi les plus élevés toutes disciplines confondues (OST 2021).

**Fig.7. Indice d'impact en infectiologie des 20 premiers producteurs**

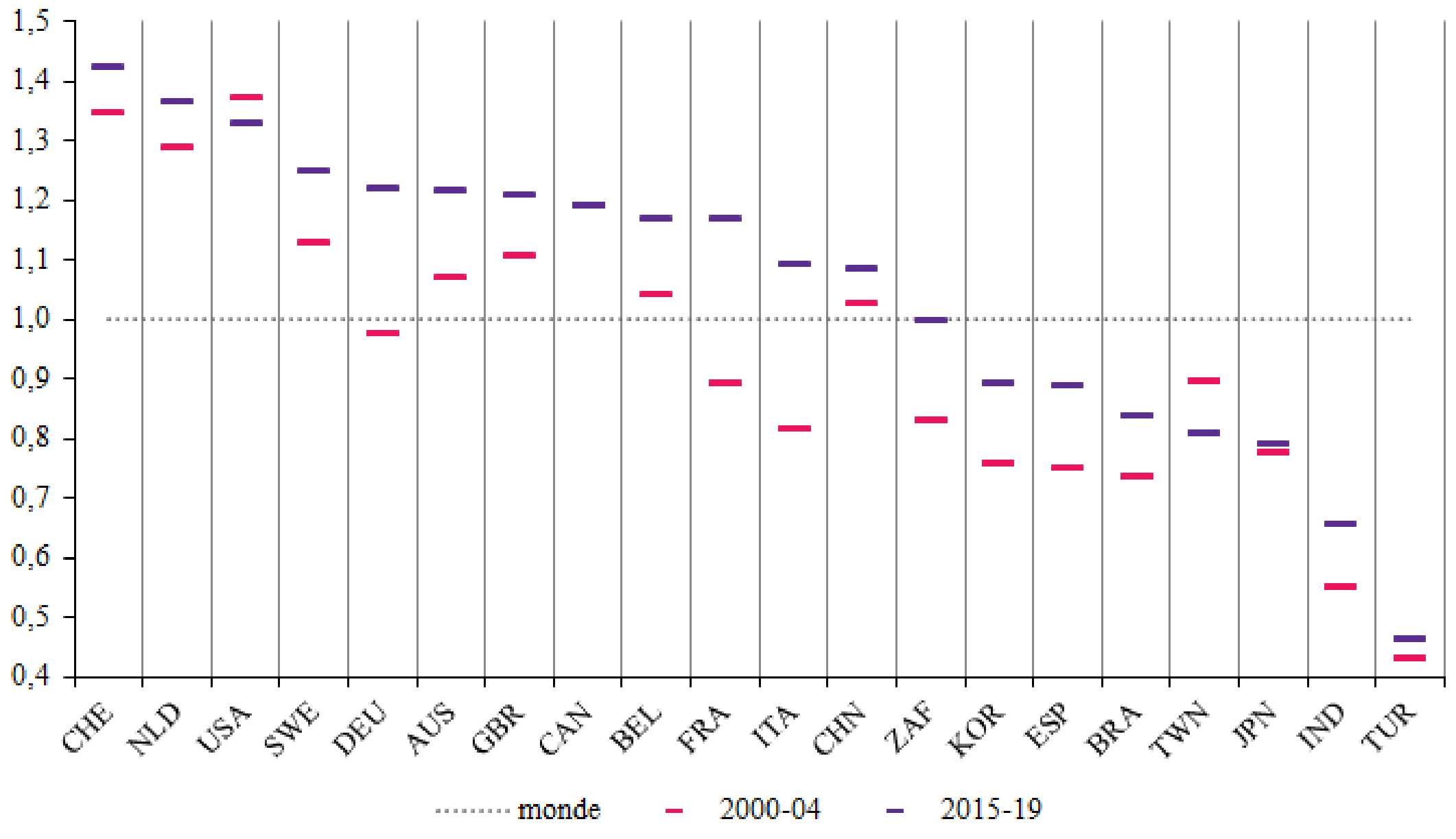

*Source : Base OST, Web of Science, calculs OST*

La France et l'Italie enregistrent une augmentation de leurs indices d'impact entre le début des années 2000 et la fin des années 2010. Leurs indices, inférieurs à la moyenne mondiale en début de période, augmentent pour atteindre respectivement 1,17 et 1,09 en 2015-19. L'impact des publications de l'Afrique du Sud augmente sensiblement sur la période et atteint la moyenne mondiale en 2015-19.

## Conclusion et approfondissements

La mobilisation de plusieurs outils d'analyse scientométrique a permis d'étudier la dynamique des publications mondiales dans le domaine de l'infectiologie de 2000 à 2020. La prise en compte de l'année 2020 a permis d'inclure les premières publications sur le Covid-19 et d'analyser l'impact de ce thème émergent sur la recherche en infectiologie en général et dans certains pays en particulier.

***Les principaux pays publiant en infectiologie ont différents profils scientifiques et sanitaires***

L'analyse du corpus mondial a souligné que les principaux pays publiant en infectiologie ne sont pas tous les mêmes que les principaux pays publiant toutes disciplines confondues. En effet, outre plusieurs pays à hauts revenus spécialisés en recherche médicale et des nouvelles puissances scientifiques telles que la Chine, l'Inde ou le Brésil, y figurent l'Afrique du Sud, Taiwan et la Turquie. La particularité de la recherche en l'infectiologie est la forte demande sociétale dans les pays où les maladies infectieuses constituent un lourd fardeau économique et social. Les pays spécialisés en infectiologie connaissent des contextes de recherche très différents : recherche médicale importante et industrie pharmaceutique développée (Suisse, Royaume Uni, Etats-Unis), infections persistantes représentant des enjeux majeurs de santé publique (Afrique du Sud, Inde, Brésil), héritage historique avec des recherches développées en lien avec des pays fortement touchés par des maladies infectieuses (France, Belgique, Espagne, Pays Bas).

***Très forte augmentation des publications en réaction à la pandémie de Covid-19***

L'analyse des publications parues en 2020 a montré que la pandémie de Covid-19 a très rapidement bouleversé le paysage mondial de la recherche en infectiologie. L'intensité et la rapidité de la production scientifique sur le sujet ont été spectaculaires. La réactivité a été particulièrement forte dans certains pays, comme l'Italie dont les publications en infectiologie ont connu une augmentation inouïe. Dans le cas de la France, la spécialisation dans le domaine du Covid n'apparaît pas différente de la spécialisation en infectiologie en général. C'est le cas pour la majorité des principaux pays publiant dans le domaine, à l'exception de l'Italie et, dans une moindre mesure, de la Chine, de la Turquie et de l'Espagne.

La pandémie a suscité une réponse inédite des communautés scientifiques, mais aussi du monde de l'édition. L'accélération de la communication s'est accompagnée d'un grand nombre de publications de faible qualité scientifique, avec des résultats non confirmés, voire erronés (Whitmore, et al. 2021). L'analyse approfondie de ces phénomènes ne pourra être menée que progressivement en fonction des données qui deviendront disponibles. L'analyse de la réponse de la communauté scientifique à la pandémie devrait aussi s'appuyer sur l'examen des co-publications internationales afin de voir dans quelle mesure les collaborations internationales ont pu contribuer à accélérer les travaux de recherche et à renforcer leur contribution scientifique. En effet, si les premières descriptions cliniques ont donné lieu à des contributions par des équipes locales (Aviv-Reuven et Rosenfeld 2021), les études plus ambitieuses tendent à mobiliser des collaborations internationales se traduisant par des co-publications internationales (Haghani, Bliemer, 2021).

***Les cartes thématiques permettent de distinguer le profil spécifique de la France au sein du domaine de l'infectiologie***

Il existe deux grands modes de constitution de corpus thématiques : à l'aide des requêtes textuelles, en langage naturel ou avec un langage contrôlé (index, thésaurus) par sélection d'éléments à partir d'une classification prédéfinie. Le recours à une nomenclature existante est plus simple mais moins performant pour circonscrire des périmètres thématiques complexes. Les requêtes textuelles en langage naturel sont lourdes à construire et sont à la fois imprécises et non exhaustives. Mais les bases citationnelles utilisées pour les analyses scientométriques ne possèdent pas de vocabulaire contrôlé. Pour pallier l'absence d'un vocabulaire contrôlé dans la base WoS, cette étude s'est appuyée sur celui de Medline pour construire l'équation de recherche. L'analyse a montré que dans le domaine de l'infectiologie le recouvrement entre la base OST-WoS et Medline est de 83%. L'absence de 17% des publications de Medline dans la base OST-WoS peuvent être multiples : des revues non indexées dans le WoS, des défauts d'appariement entre les deux bases, la fraicheur de la mise à jour, etc. Cette question n'a pas été étudiée et pourrait constituer un approfondissement en vue d'une amélioration de la méthode.

Les cartes thématiques construites à partir des cooccurrences de mots clés dans les publications ont permis d'identifier 7 clusters au sein du corpus global des publications en infectiologie. L'analyse a ensuite mis en évidence les thématiques qui distinguent les publications françaises. Elles sont plus concentrées sur deux des clusters du corpus mondial, Infections opportunistes et secondaires d'une part et Zoonoses, maladies à transmission vectorielle et maladies tropicales d'autre part. Symétriquement, les publications françaises traitent relativement moins que le monde les thématiques Mécanismes fondamentaux de l'infection et interactions hôte-pathogène d'une part et Maladies infectieuses à transmission directe et santé publique d'autre part. En 2020, la France est apparue presque aussi spécialisée sur le thème du Covid-19 que sur l'infectiologie en général, ce qui ne suggère pas une orientation forte des recherches en faveur de ce thème. Le profil thématique de la France au sein de l'infectiologie demandera à être confirmé sur une période d'observation plus longue. Avec plus de recul, il pourrait en outre être possible d'analyser d'éventuels liens entre les difficultés de coordination identifiées au sein du système de recherche dans les débuts de la pandémie et les publications scientifiques issues des travaux sur le Covid-19. Dans cette perspective plus générale, il sera aussi utile de confronter les données de publications avec celles des essais cliniques.

# Références

# Annexes

**Fig.A1. Cartographie temporelle des liens du mot-clé Covid-19 avec le reste du corpus**

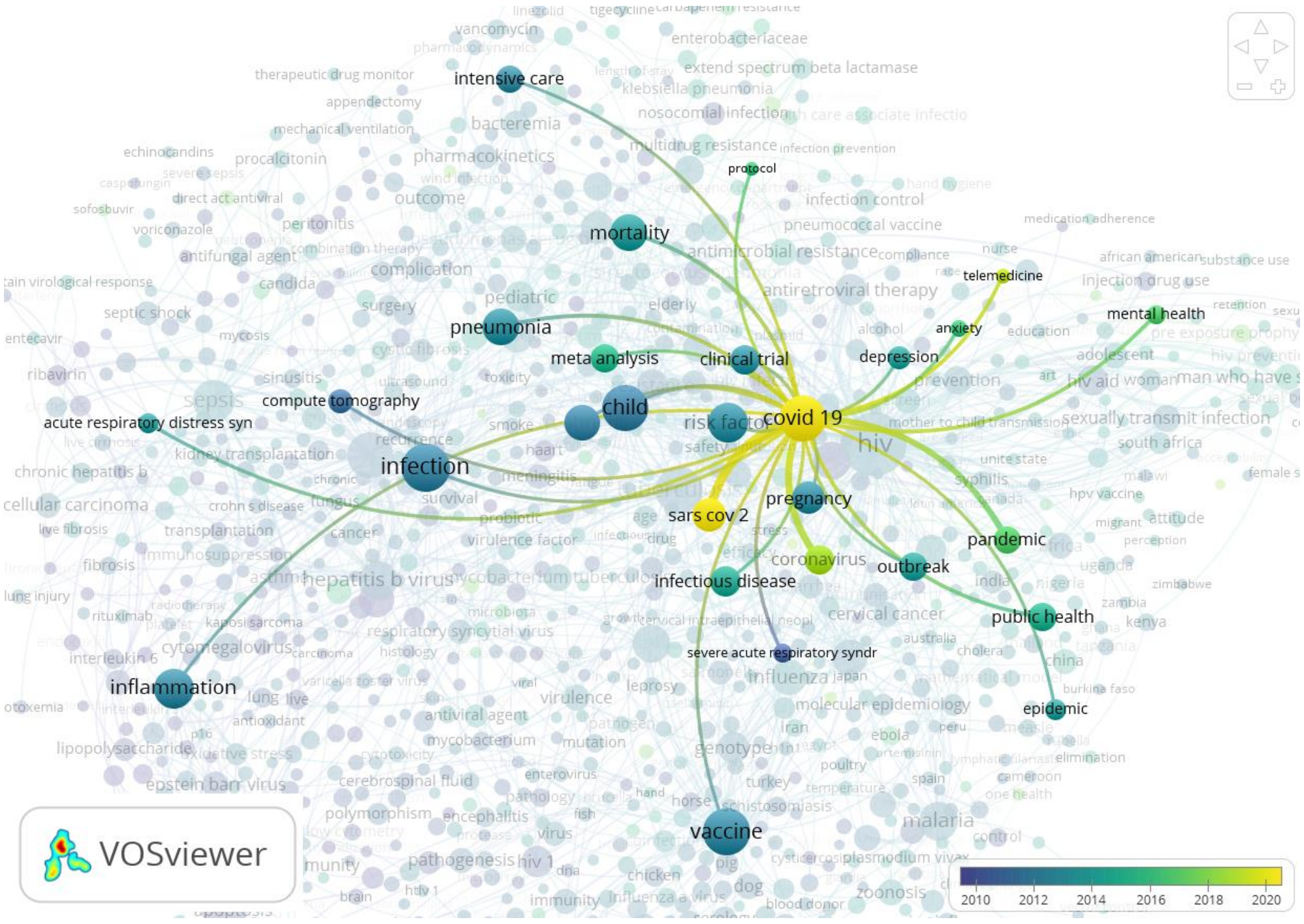

La figure A1 montre les liens de cooccurrence du terme « Covid-19 » avec les autres termes. Le dégradé de couleur permet de visualiser la concentration des termes en fonction des années. Ainsi, plus la couleur de la bulle est jaune, plus le terme et la thématique liée à ce dernier sont récents. Dans la littérature, on parle communément de « research hotspots » pour désigner les recherches qui sont d'actualité et celle qui relèvent des périodes anciennes.

Comme on peut le constater sur la carte, sans surprise, la recherche sur le Covid-19 est fortement concentrée sur les deux dernières années, contrairement aux recherches sur le SARS dont la couleur apparait en violet foncé. L'intérêt d'une telle représentation dans le cas du Covid-19 est qu'elle permet de visualiser, d'une part, les thématiques qui y sont liées (analyser les différents prismes sous lesquels la question est traitée), et d'autre part, étudier le caractère nouveau ou ancien de celles-ci.

On constate que le terme « Covid-19 » est lié à l'ensemble des clusters du corpus infectiologie. Cela témoigne de l'étendu des recherches engagées sur le sujet au sein de toutes les communautés de recherche en infectiologie. Néanmoins, force est de constater que le cluster 4 « Maladies infectieuses à transmission directe et santé publique » (où se trouve le terme « Covid-19 » par ailleurs) représente une part importante des liens de cooccurrence. Au sein de ce cluster, le terme Covid-19 est associé à plusieurs termes désignant des « research hotspots » à l'instar de la « télémédecine », « santé mentale » ou encore « pandémie ». De même, le terme « Covid-19 » est associé avec des termes comme « santé publique », « maladies infectieuses » et « épidémie », dans les publications scientifiques relevant du cluster 1 qui regroupe les recherches sur les « Zoonoses, maladies à transmission vectorielle, maladies tropicales ».

À l'opposé, le terme « Covid-19 » est associé avec des termes dont la présence est relativement plus forte dans les années avant 2012. Il s'agit par exemple des termes « computer tomography », « infection », « inflammation », « child » et « treatment » du cluster 3 et « vaccin » du cluster 1.

## Tableau A1 : mots-clés Covid

| **Mots clefs** | **Mots cles utilisés** |
|---|---|
| 2019 NCOV | 2019 ncov |
| 2019 Novel Coronavirus | coronavirus |
| coronavirus (2019-nCoV) outbreak% | coronavirus |
| coronavirus (CoV) | coronavirus |
| coronavirus 2 | coronavirus |
| coronavirus 2019 | coronavirus |
| coronavirus 2019 disease | coronavirus |
| coronavirus 2019-nCoV | coronavirus |
| coronavirus crisis | coronavirus |
| coronavirus disease 19 | coronavirus |
| coronavirus disease 2 | coronavirus |
| coronavirus disease 2019 | coronavirus |
| coronavirus epidem% | coronavirus |
| coronavirus identified in 2019 | coronavirus |
| coronavirus mortality | coronavirus |
| coronavirus outbreak% | coronavirus |
| coronavirus pandemic | coronavirus |
| coronavirus pneumonia infection | coronavirus |
| coronavirus S protein | coronavirus |
| coronavirus SARS | coronavirus |
| coronavirus SARS-CoV-2 | coronavirus |
| coronavirus spike | coronavirus |
| coronavirus vaccine% | coronavirus |
| coronavirus-2019 pneumonia | coronavirus |
| coronavirus-infected pneumonia | coronavirus |
| COVID | Covid |
| SARS Coronavirus 2 | coronavirus |
| SARS-CoV-2 | sars cov 2 |
| Severe Acute Respiratory Syndrome Coronavirus 2 | coronavirus |
| Wuhan Coronavirus | coronavirus |
| Wuhan Seafood Market Pneumonia Virus | wuhan seafood market pneumonia virus |
| SARS-CoV2 | sars cov2 |